\documentclass[prd,twocolumn,superscriptaddress,preprintnumbers,balancelastpage,nofootinbib]{revtex4-1}
\usepackage{graphicx}  
\usepackage{dcolumn}   
\usepackage{bm}        
\usepackage{amssymb, amsmath}
\usepackage{slashed}   
\usepackage{lipsum}
\usepackage[usenames,dvipsnames,svgnames,table]{xcolor}
\usepackage{comment}
\usepackage[utf8]{inputenc}
\renewcommand{\thefootnote}{\fnsymbol{footnote}}

\newcommand{\GeV}{{\rm GeV}}

\newcommand{\ov}{\overline}

\RequirePackage{xspace}
\def\Bbar    {\kern 0.18em\overline{\kern -0.18em B}{}\xspace}
\def\Bb      {\ensuremath{\Bbar}\xspace}
\def\Kbar    {\kern 0.18em\overline{\kern -0.18em K}{}\xspace}

\renewcommand{\thefootnote}{\#\arabic{footnote}}
\definecolor{BlueViolet}{rgb}{0.2, 0.00, 0.7}
\definecolor{Blue}{rgb}{0.15, 0.00, 0.9}
\definecolor{lightblue}{rgb}{0.15, 0.35, 0.95}
\definecolor{kitgreen}{rgb}{0,
0.58823 
, 0.50980 
}
\usepackage[
colorlinks=true, linkcolor=lightblue,citecolor=lightblue,urlcolor=kitgreen]{hyperref} 

\begin{document}
\widetext
\preprint{P3H--23--010, TTP23--05}

\title{Coup de grace to the charged Higgs solution of $P_5^\prime$ and $R_{D^{(*)}}$ discrepancies}

\author{Syuhei Iguro}
\email{igurosyuhei@gmail.com}
\affiliation{Institute for Theoretical Particle Physics (TTP), Karlsruhe Institute of Technology (KIT),
Engesserstra{\ss}e 7, 76131 Karlsruhe, Germany}
\affiliation{ Institute for Astroparticle Physics (IAP),
Karlsruhe Institute of Technology (KIT), 
Hermann-von-Helmholtz-Platz 1, 76344 Eggenstein-Leopoldshafen, Germany}

\begin{abstract}
\noindent
We consider a general two Higgs doublet model which can simultaneously solve discrepancies in neutral B meson decay ($b\to s\ell \ov \ell$ distribution) and charged B meson decay ($b\to c\tau\ov\nu$) with a charged Higgs.
The model contains two additional neutral scalars at the same mass scale and predicts distinctive signals at the LHC.
Based on the recent same-sign top search by the ATLAS collaboration, we found the constraint on the scalar mass spectrum.
To probe the remaining mass window, we propose a novel $cg\to t\tau\ov\tau$ process at the LHC.  \\
--------------------------------------------------------------------------------------------------------------------------------
\end{abstract}

\maketitle

\renewcommand{\thefootnote}{\#\arabic{footnote}}
\setcounter{footnote}{0}

\section{Introduction}
\label{Sec:intro}
The current flavor anomalies in B meson decays $\it{e.g.}$ deviations in angular distribution in $b \to s \mu\ov\mu$ processes, so-called $P_5^\prime$ \cite{LHCb:2013zuf,LHCb:2013tgx,LHCb:2013ghj,CMS:2013mkz,LHCb:2014cxe,LHCb:2015wdu,LHCb:2015svh,LHCb:2020lmf,CMS:2020oqb,LHCb:2020gog,LHCb:2021xxq}\footnote{Different from lepton flavor universality ratio $R_{K^{(*)}}={\rm{BR}}(B\to K^{(*)} \mu\ov\mu)/{\rm{BR}}(B\to K^{(*)} e\ov{e})$, there is sizable hadronic parameter dependence.
For instance sizable charm hadronic contributions would also explain the deviation, see Refs,\,\cite{Ciuchini:2015qxb,Ciuchini:2022wbq} for instance.
On the other hand, the tension between measured BR($B_s\to \phi \mu\ov\mu$) \cite{LHCb:2021zwz}, BR($\Lambda_b\to\Lambda \mu\ov\mu$) \cite{LHCb:2015tgy} and BR($B \to K^{(*)}\mu\ov\mu$) \cite{LHCb:2014cxe} and the SM predictions \cite{Ball:2004rg,Horgan:2013pva,Horgan:2015vla,Bharucha:2015bzk,Detmold:2012vy,Bobeth:2011gi,Bobeth:2011nj} can be relaxed with the vector contribution.}, and lepton flavor violation of $\Bb \to D^{(*)} \tau\ov\nu$ \cite{Lees:2012xj,Lees:2013uzd,Huschle:2015rga,Hirose:2016wfn,Hirose:2017dxl,Belle:2019gij,Belle:2019rba,Aaij:2015yra,Aaij:2017uff,Aaij:2017deq,Bordone:2019guc,Iguro:2020cpg,Bernlochner:2022ywh,HFLAV:2022pwe} can be solved with a light charged scalar ($H^+$) from a generic two Higgs doublet model (G2HDM) \cite{Iguro:2018qzf,Kumar:2022rcf}\footnote{The possibility was originally pointed out in Ref.\,\cite{Iguro:2018qzf}, and recently revisited in Ref.\,\cite{Kumar:2022rcf}.
It is noted that thanks to the relaxed constraint from $B_c\to\tau\ov\nu$ \cite{Alonso:2016oyd,Celis:2016azn,Blanke:2018yud,Aebischer:2021ilm} and the experimental shift, $H^+$ can now explain $R_{D^{(*)}}={\rm{BR}}(\Bb\to D^{(*)} \tau\ov\nu)/{\rm{BR}}(\Bb\to D^{(*)} \ell\ov\nu)$ within $1\,\sigma$ \cite{Iguro:2022FCCP}.
For the individual explanation, see, Refs.\,\cite{Hu:2016gpe,Arnan:2017lxi,Arhrib:2017yby,Li:2018rax,Crivellin:2019dun} for $b\to s \ell\ov\ell$ and Refs.\,\cite{Crivellin:2012ye,Crivellin:2013wna,Cline:2015lqp,Crivellin:2015hha,Lee:2017kbi,Iguro:2017ysu,Martinez:2018ynq,Fraser:2018aqj,Athron:2021auq,Iguro:2022uzz,Blanke:2022pjy,Fedele:2022iib} for $R_{D^{(*)}}$. 
}.
Although a significant deviation in the lepton flavor universality test in $b\to s \ell\ov\ell$ transition where $\ell=e,\,\mu$ has disappeared in the recent LHCb measurement \cite{LHCb:2022qnv,LHCb:2022zom} thanks to the improved electron tagging method.
Furthermore, the deviation in $B_s\to \mu\ov\mu$ has gone \cite{CMS:2022dbz} and, consequently the explicit priority of the vector$-$axial vector (V$-$A) like interaction no longer exists \cite{Ciuchini:2022wbq}.
Those recent changes brought charged Higgs solution back into the game and makes it more appealing.
There days, due to the disappearance of $R_{K^{(*)}}$ puzzle, there is a psychological tone down for the B anomalies, though, it is a fact that there are still about $3\sim4\,\sigma$ discrepancies in $b\to s \ell\ov\ell$ and $b\to c\tau\ov\nu$ processes.  

Interestingly a successful charm penguin contribution to the flavor universal vector operator of $b\to s \ell\ov\ell$ and tree level $b\to c\tau\ov\nu$ transition are both controlled by the common $\ov b_L c_R H^-$ interaction where the corresponding Yukawa coupling is denoted as $\rho_u^{tc}$.
In the G2HDM, the coupling $\rho_u^{tc}$ induces $\ov t_L  c_R\phi$ interaction where $\phi=H,\,A$ denotes additional neutral scalars which are SU(2)$_L$ partners of the charged Higgs.
It is noted that the additional doublet with sizable $\rho_u^{tc}$ is discussed with the spontaneous CP violating scenario \cite{Lee:1973iz,Nierste:2019fbx} and the electroweak baryogenesis \cite{Fuyuto:2017ewj}.\footnote{They used the closed time path formalism \cite{Riotto:1998bt} to evaluate the produced baryon number.} 

The available mass range of the  charged scalar for the simultaneous explanation is bounded from the above based on the $\tau\ov\nu$ resonance searches at the LHC \cite{CMS:2018fza} as $m_{H^+}\le400\,\GeV$ \cite{Iguro:2018fni}. 
Although in Ref.\,\cite{Blanke:2022pjy} we theoretically showed that the $b+\tau\ov\nu$ resonance search is a powerful tool to probe the remaining parameters, the corresponding experimental search has not been performed.

Different from recent studies which mainly focus on the charged scalar collider phenomenology in light of deviations in B meson decays \cite{Iguro:2022uzz,Blanke:2022pjy,Desai:2022zig}, we consider the collider signal of additional neutral scalars.
Although connection between sizable $\rho_u^{tc}$ and neutral scalars mediated multi-top final states at the LHC has been discussed in Refs.\,\cite{Iguro:2017ysu,Gori:2017tvg,Kohda:2017fkn,Iguro:2018fni,Hou:2018zmg,Hou:2020tnc,Hou:2020chc},\footnote{See, also Refs.\,\cite{Hall:1981bc,Hou:1997pm,Atwood:1996vj,Altunkaynak:2015twa} for the earlier works to probe $\rho_u^{tc}$ in a flavor changing top decay.}
last summer, the ATLAS collaboration reported the game changing result \cite{ATLAS:2022xpz}.
They searched for the G2HDM in top-associated processes and directly set the upper limit on $\rho_u^{tc}$.
In this letter, we reinterpret the constraint in light of the simultaneous explanation and propose an additional process to cover the remaining parameter space thorough the neutral scalars.
Thanks to the electroweak precision data even after the controversial CDF result \cite{CDF:2022hxs}, the mass of those additional scalars ($m_\phi$) should be similar to $m_{H^+}$ up to $\mathcal{O}(v)$ where $v=246\,\GeV$ denotes the vacuum expectation value.
Therefore it would be natural to consider the LHC phenomenology to fully probe the interesting parameter space.
  
The outline of the letter is given as follows.
In Sec.\,\ref{Sec:Model} we introduce the model setup and explain the relevant parameters.
The favored region and upper limit on the additional scalars are summarized in Sec.\,\ref{Sec:param}.  
In Sec.\,\ref{Sec:ETP} we investigate the model prediction of top-associated processes.
Summary and discussion will be given in Sec.\,\ref{Sec:Summary}.

\section{Model setup}
\label{Sec:Model}
We consider a two Higgs doublet model (2HDM) where an additional scalar doublet is introduced to the SM.
The general scalar potential of the model is given as
\begin{align}
  \rm{V}(H_1,\,&H_2)=M_{11}^2 H_1^\dagger H_1+M_{22}^2 H_2^\dagger H_2-\left(M_{12}^2H_1^\dagger H_2+{\rm h.c.}
  \right)\nonumber\\
&+\frac{\lambda_1}{2}(H_1^\dagger H_1)^2+\frac{\lambda_2}{2}(H_2^\dagger H_2)^2+\lambda_3(H_1^\dagger H_1)(H_2^\dagger H_2)\nonumber\\
&+\lambda_4 (H_1^\dagger H_2)(H_2^\dagger H_1) 
+\frac{\lambda_5}{2}(H_1^\dagger H_2)^2\nonumber\\
&+\left\{
\lambda_6 (H_1^\dagger H_1)+\lambda_7 (H_2^\dagger H_2)\right\} (H_1^\dagger H_2)+{\rm h.c.}.
   \label{Eq:potential}
\end{align}
Here, we  work in the {\em Higgs basis} where only one doublet takes the VEV \cite{Georgi:1978ri,Donoghue:1978cj}:
\begin{align}
H_1 = \begin{pmatrix} G^+ \\ \frac{1}{\sqrt{2}} (v+h+iG^0) \end{pmatrix}, \,
H_2 = \begin{pmatrix} H^+ \\ \frac{1}{\sqrt{2}} (H+iA) \end{pmatrix},
 \label{Eq:basis}
\end{align}
where $G^{+}$ and $G^{0}$ denotes the NG bosons.
It is noted that alignment where the SM h lives in $H_1$ is considered to avoid the constraint from $t\to c h$ \cite{CMS:2021hug,CMS:2021gfa,ATLAS:2023mcc}.
For simplicity, we further assume the CP-conserving scalar potential and then one can define the CP-even and -odd scalar mass eigenstates.
The SM-like Higgs is $h$ and $H$ and $A$ correspond to additional the CP-even and -odd neutral scalars.
Masses differences among additional scalars are given as,
\begin{align}
  m_H^2& =m_A^2+\lambda_5 v^2,\,\,\, m_{H^+}^2 = m_A^2-\frac{\lambda_4-\lambda_5}{2} v^2.
  \label{Eq:Higgs_spectrum}
\end{align}
It is noted that other potential couplings does not affect the following discussion.

When the both doublets couple to all fermions, the Higgs bosons have flavor violating interactions in general.
In this letter we take the bottom-up approach and introduce the interaction Lagrangian of the heavy scalars relevant to $b\to s \ell\ov\ell$ and $b\to c\tau\ov\nu$,
\begin{align}
{\cal L}_{int}&=
\,\rho_u^{tc} \frac{H+iA}{\sqrt{2}} (\overline{t} P_R c)
+ \rho_e^{\tau\tau}\frac{H-iA}{\sqrt{2}} (\overline{\tau} P_R \tau)\nonumber\\
& + V^*_{td_i} \rho_u^{tc} H^- (\overline{d_i} P_R c)
- \rho_e^{\tau\tau} H^- (\overline{\tau} P_L \nu_{\tau})   +{\rm{h.c.}},
\label{Eq:G2HDM_Yukawa}
\end{align}
where $P_{L/R}=(1\mp\gamma_5)/2$ and $V$ are a chirality projection operator and Cabbibo-Kobayashi-Maskawa matrix \cite{Cabibbo:1963yz,Kobayashi:1973fv}, respectively. 
The neutral scalar interaction and the charged scalar interaction are related by the SU(2)$_{\rm{L}}$ rotation.
We assume that other Yukawa coupling to be small ($\ll \mathcal{O}(10^{-2})$) for simplicity.
For the more detailed phenomenological analysis with other Yukawa couplings, see Refs.\,\cite{Crivellin:2013wna,Iguro:2017ysu,Iguro:2019zlc}.
We will also discuss this point in Sec.\,\ref{Sec:Summary}.

For the later convenience we show the approximate formulae for the partial decay width,
\begin{align}
    \Gamma(\phi \to \tau\ov\tau)\simeq\frac{|\rho_e^{\tau\tau}|^2}{16\pi}m_\phi,\,
    \Gamma(\phi  \to t c)\simeq
    \frac{3|\rho_u^{tc}|^2 m_\phi}{16\pi} \beta^2(m_\phi),
\end{align}
where $\Gamma(\phi  \to t c)=\Gamma(\phi  \to t \ov{c})+\Gamma(\phi  \to \ov{t} c)$ and $\beta(m_\phi )= \left(1-\frac{m_t^2}{m_\phi^2}\right)$ are defined.\footnote{In this letter we neglect light fermion masses, though, one can trivially include the effect.}

\section{Summary of the available parameter region} 
\label{Sec:param}
\begin{figure}[t]
\begin{center}
\includegraphics[scale=0.45]{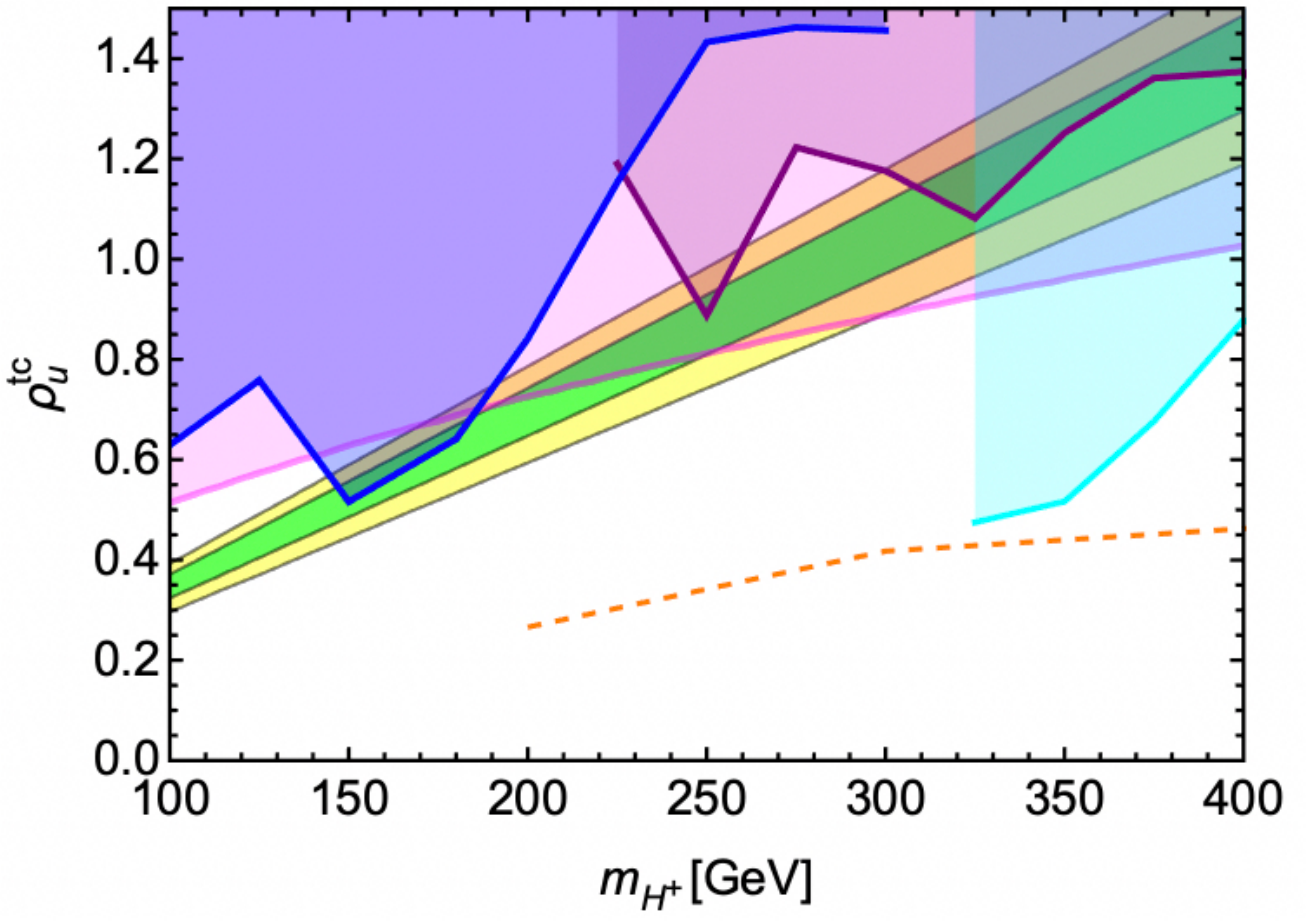}
\caption{
\label{Fig:ptcvsmHp}
The favored region of $C_9^\ell$ is shown in green ($1\,\sigma$) and yellow ($2\,\sigma$) on the $\rho_u^{tc}$ vs. $m_{H^+}$ plane.
$B_s-\ov{B_s}$ mixing constraint excludes the magenta region.
Cyan, purple, blue regions are excluded by low mass di-jet resonance searches.
The orange dashed line corresponds to the upper limit from the same-sign top search adopted from Ref.\,\cite{ATLAS:2022xpz} assuming $m_{H^+}=m_H$.
See the main text for further detail.
} 
\end{center}
\end{figure}

First we consider the charged Higgs contribution to flavor universal $b\to s\ell\ov\ell$.
Since the coupling dependence is different among $b\to s\ell\ov\ell$  (induced by the charm penguin $\propto|\rho_u^{tc}|^2$)
and the most constraining flavor process, $B_s-\ov{B_s}$ mixing (charged Higgs box $\propto|\rho_u^{tc}|^4$), we can set an upper limit on the charged Higgs mass \cite{Iguro:2018qzf,Kumar:2022rcf}.
The relevant Hamiltonian for $b\to s\ell\ov\ell$ in our model is given as
\begin{align}
{\mathcal{H}}_{\rm{eff}}&=-\frac{\alpha G_F}{\sqrt{2}\pi}V_{tb}V_{ts}^*
C_{9}(\overline{s}\gamma^{\mu}P_{L} b)(\overline{l}\gamma_\mu l)+\rm{h.c.},\label{Eq:effH}
\end{align}
where $l=e,\,\mu$ and $\tau$.
We note that contribution from $Z$ penguin is small enough to neglect.
We follow the prescription in 
Ref.\,\cite{Kumar:2022rcf} and use the following numerical formula,
\begin{align}
    C_9^l(\mu_b)\simeq -0.95\left(\frac{|\rho_u^{tc}|}{0.7}\right)^2\left(\frac{200\,\GeV}{m_{H^+}}\right)^2.
    \label{Eq:bsll_simplified}
\end{align}
This should be compared with the recent global fit to $b\to s\ell\ov\ell$ data of $C_9^l(\mu_b)=-0.95\pm0.13$ \cite{Hurth:2022lnw}.\footnote{This fit does not include $B_s\to\mu\ov\mu$ and lepton flavor universality observables {\it{e.g.}} $R_{K^{(*)}}$.
Since $C_9$ operator does not contribute to $B_s\to\mu\ov\mu$, the result will be unchanged, though.
The similar result is also reported in Ref.\,\cite{Ciuchini:2022wbq}.}
In Fig.\,\ref{Fig:ptcvsmHp}, we show $1\,(2)\,\sigma$ favored region in green (yellow) on the $m_{H^+}$ vs. $\rho_u^{tc}$ plane. 
Since we also has the upper limit on the mass as $m_{H^+}\le 400\,\GeV$ and the lower limit form the LEP experiment \cite{ALEPH:2013htx}, we focus on $100\,\GeV\le m_{H^+}\le 400\,\GeV$.
As mentioned above $B_s$ meson mixing puts the most stringent flavor constraint \cite{DiLuzio:2019jyq} which is shown in magenta.

In this mass region, di-jet resonance searches at the LHC are able to set the upper limit on $\rho_u^{tc}$ \cite{Iguro:2022uzz}.
We overlay the constraint from the (bottom flavored) di-jet searches in blue \cite{CMS:2017dcz}, purple \cite{ATLAS:2019itm} and cyan \cite{CMS:2018kcg} where BR($H^+\to \ov{b}c$)=1 is assumed.
It is noted that as we will see soon later, we need a hierarchy of $|\rho_u^{tc}|\gg|\rho_e^{\tau\tau}|$ for the simultaneous explanation.
As a result $H^+\to \ov{b}c$ is the dominant decay mode in the minimal set up of Eq.\,(\ref{Eq:G2HDM_Yukawa}) and hence the exclusion discussed above is unaffected.\footnote{The stau search constraint \cite{ATLAS:2019gti,CMS:2022rqk} on the charged Higgs is very weak due to BR($H^+\to \ov{b} c)\simeq1$.
See, Fig.\,4 of Ref.\,\cite{Iguro:2022tmr}.}
We see that di-jet constraints touch the interesting parameter region.
Run\,2 full data would be possible to improve the constraint further.

\begin{figure}[t]
\begin{center}
\includegraphics[scale=0.23]{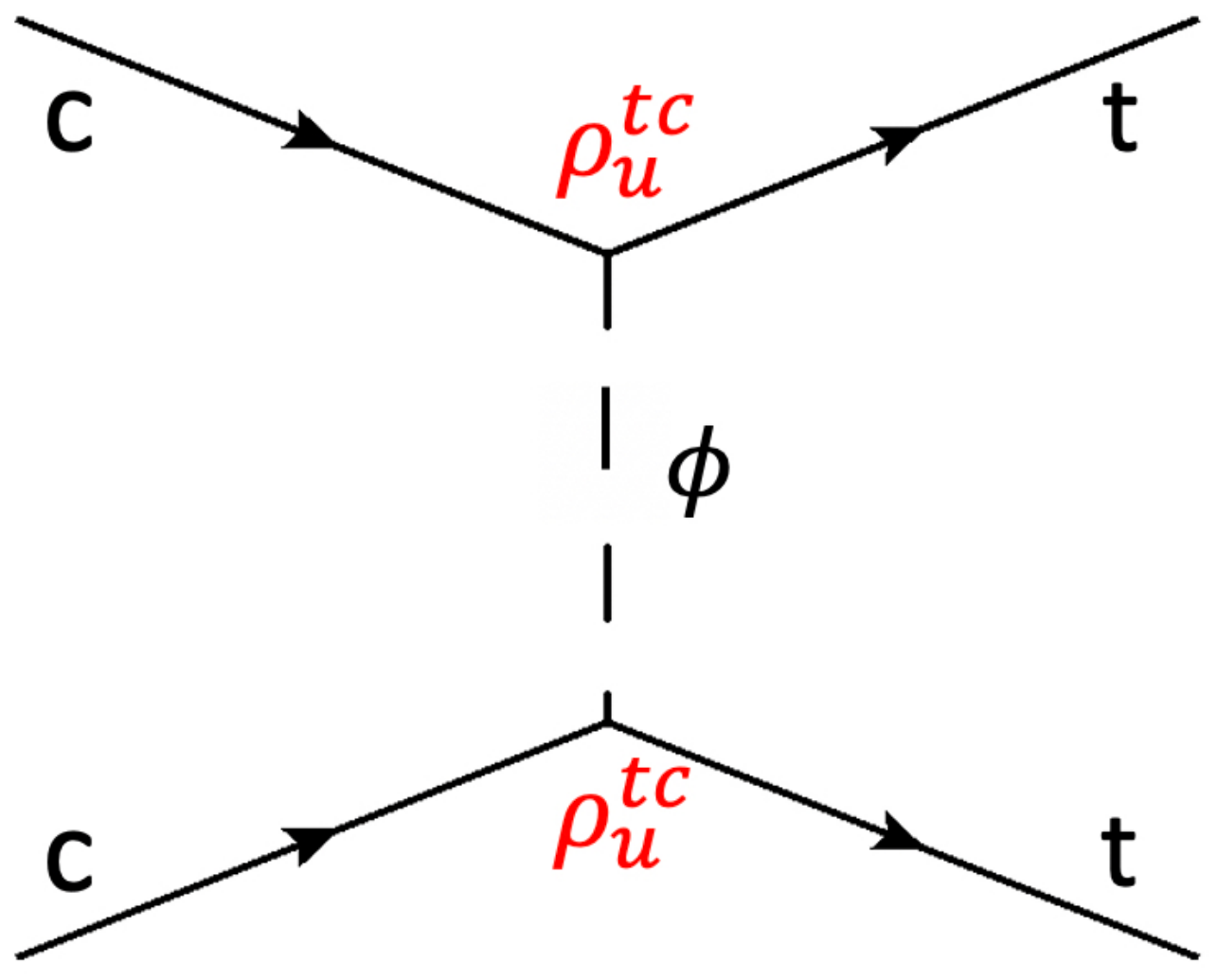}~~
\includegraphics[scale=0.23]{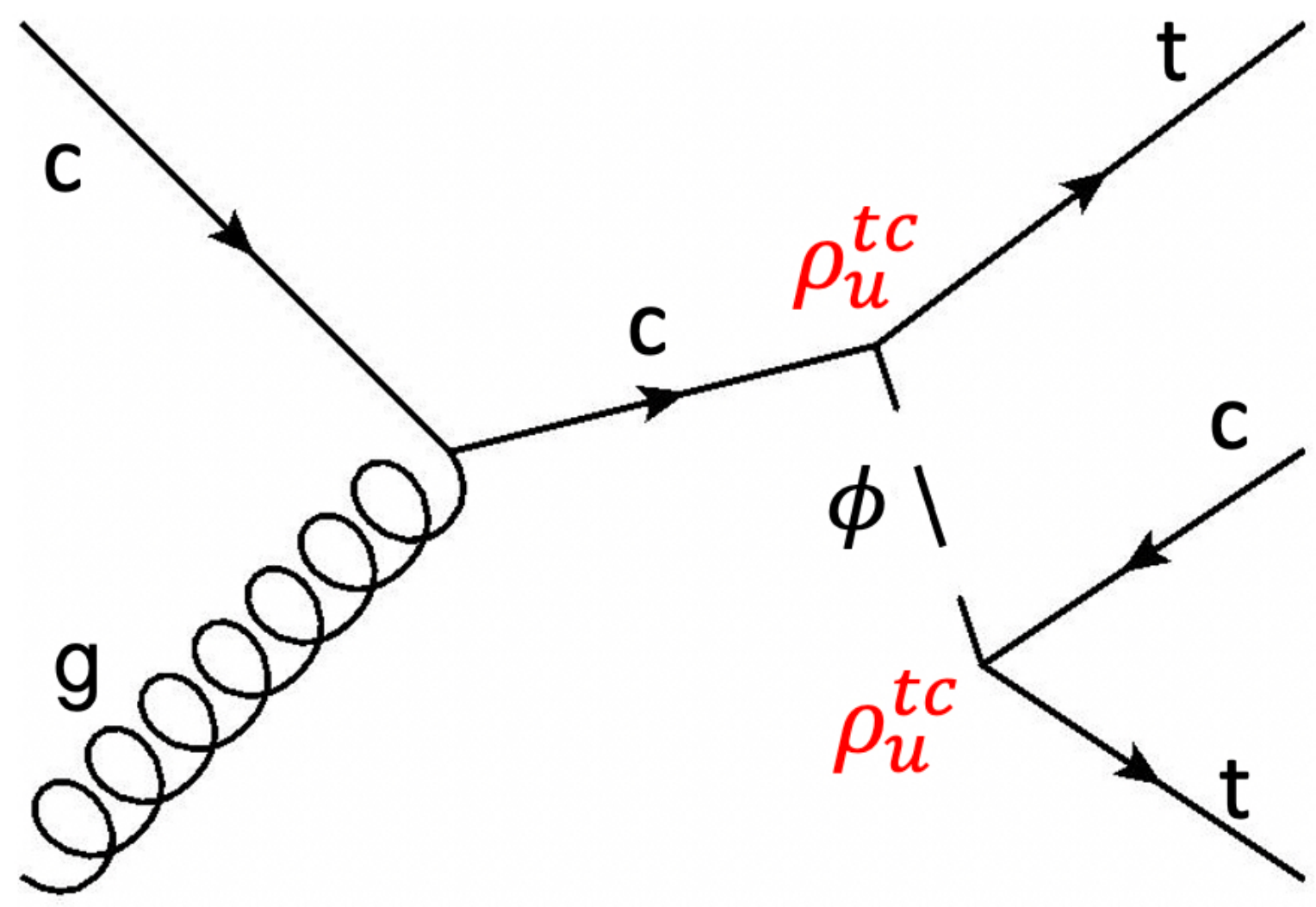}
\caption{
\label{Fig:SSt}
The representative diagrams for the same-sign top final state at the LHC.
In the numerical evaluation we include the charge conjugated processes also.
The dominant contribution comes from the right diagram.
} 
\vspace{-.45cm}
\end{center}
\end{figure}

We move onto the explanation of the $R_{D^{(*)}}$ discrepancy.
The relevant interaction Hamiltonian is given as
\begin{align}
 {\mathcal H}_{\rm{eff}}= 
 2 \sqrt 2 G_F V_{cb}  C_{S_L}^\tau(\overline{c} P_L b)(\overline{\tau} P_L \nu_{\tau}) .
 \label{Eq:Hamiltonian_RD}
\end{align} 
The charged Higgs contribution including renormalization group running corrections \cite{Alonso:2013hga, Jenkins:2013wua,Gonzalez-Alonso:2017iyc,Aebischer:2017gaw}, is approximately given as
\begin{align}
    |C_{S_L}^\tau(\mu_b)|\simeq 0.83\left(\frac{|\rho_u^{tc*}\rho_e^{\tau\tau}|}{0.03}\right)\left(\frac{200\,\GeV}{m_{H^+}}\right)^2.
    \label{Eq:RD_simplified}
\end{align}
Adopting the analytic formulae of $R_{D^{(*)}}$ in Ref.\,\cite{Iguro:2022uzz}\footnote{Those analytic formulae used in Ref.\,\cite{Iguro:2022uzz} are consistent with the recent result \cite{Iguro:2022yzr} within the uncertainty.} latest $1\,\sigma$ explanation is realized with $0.68\lesssim|C_{S_L}^\tau(\mu_b)|\lesssim 1.13$.\footnote{To fit the $R_{D^{(*)}}$ data $\rho_u^{tc*}\rho_e^{\tau\tau}$ needs to have a complex phase, however, this does not change the following discussion.}
By combining Eqs.\,(\ref{Eq:bsll_simplified},\,\ref{Eq:RD_simplified}), one can see that the simultaneous explanation requires the large magnitude difference in $\rho_u^{tc}$ and $\rho_e^{\tau\tau}$.

So far we focused on the charged Higgs phenomenology, however, neutral scalar mass spectrum is constrained with the LHC data and electroweak precision observables.  
The last summer the ATLAS collaboration reported the result of the G2HDM search in top-associated processes \cite{ATLAS:2022xpz} for $m_\phi\ge\, 200\,\GeV$.\footnote{To adopt the experimental data and extend the constraint down to $m_\phi\simeq m_t$, detailed distribution data is necessary.
Although this data is not available in Ref.\,\cite{ATLAS:2022xpz} and thus beyond scope of this letter.}
The relevant signal events include the same-sign top quarks.
In Fig.\,\ref{Fig:ptcvsmHp}, the constraint directly taken from Ref.\,\cite{ATLAS:2022xpz} is shown in the orange dashed line assuming $m_H=m_{H^+}$.\footnote{In this analysis they only considered $H$ to be present and ignore $A$ for simplicity. 
If there is a mild mass difference of $\mathcal{O}(10)\,\GeV$, the constraint will be more stringent by a factor of $\sqrt{2}$. 
}
It is observed that this same-sign top search would exclude the $b\to s\ell\ov\ell$ explanation for $m_\phi\ge\, 200\GeV$.
Although there is a loophole in this same-sign top bound.
There are two-types of the contributing Feynman diagrams, namely t-channel (left) and s-channel (right) as shown in Fig.\,\ref{Fig:SSt}.
In both diagrams, due to the different CP nature of $H$ and $A$, the amplitude cancels in the mass degenerate limit.
The destructive interference for the dominant s-channel approximately happens up to the width difference \cite{Hou:2018zmg}.
For the simultaneous explanation, $\rho_u^{tc}$ needs to be as large as 0.7 (0.8) for $m_{H^+}=200\,(250)\,\GeV$ and hence the total width of $\Gamma_{\phi}=0.8\,(3.5)\,\GeV$ is predicted.
This indicates that $|\lambda_5|\le{\mathcal{O}}(10^{-2})$ is necessary for the simultaneous explanation with $m_\phi \ge 200\,\GeV$.
To simplify the analysis and evade the constraint we set $m_A=m_H$ in the following.

\begin{figure}[t]
\begin{center}
\includegraphics[scale=0.33]{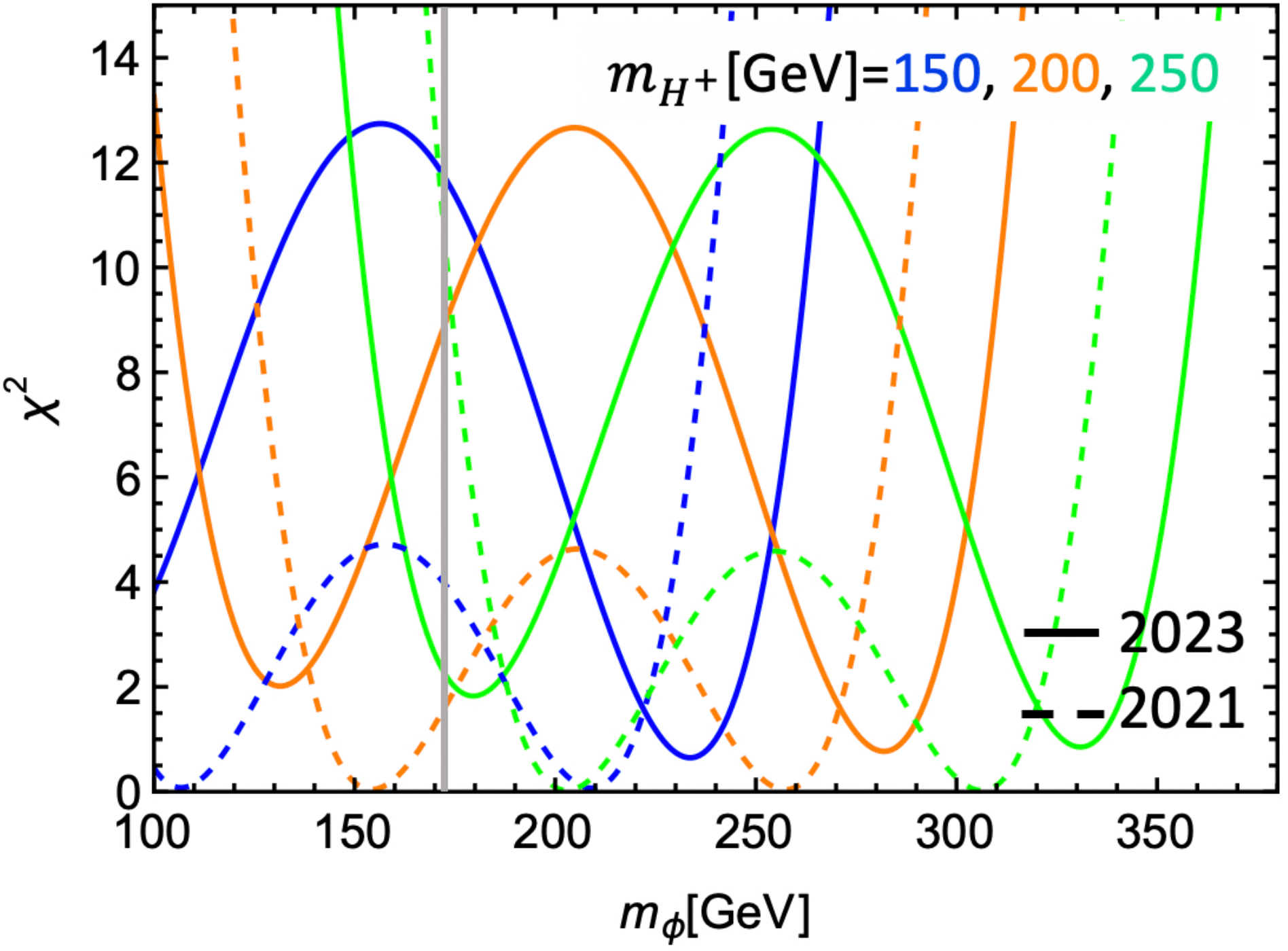}~~
\caption{
\label{Fig:chi2_ST}
The $\chi^2$ based on $S$ and $T$ parameters before (dashed) and after (solid) the recent CDF result is shown as a function of $m_\phi$.
For blue, orange and green lines, $m_{H^+}=150,\,200,\,250\,$GeV are fixed.
The gray vertical line corresponds to $m_\phi=m_t$.
}
\vspace{-.45cm}
\end{center}
\end{figure}

On the other hand, additional neutral scalars dominantly decay to $\tau\ov\tau$ for $m_\phi\le m_t$. 
In that case, the electroweak pair production of neutral scalars results in multiple $\tau$ final state.
Such a region is studied in Ref.\,\cite{Chun:2015hsa} and even only with Run\,1 data \cite{ATLAS:2014ikz} we can exclude our scenario of $m_\phi\le m_t$. 
Furthermore do not have an explicit new physics signal with the Run 2 full data \cite{CMS:2021cox,ATLAS:2022nrb} and hence the exclusion is robust.

Besides, electroweak precision observables are helpful to further constrain the mass spectrum.
We consider $S$ and $T$ parameter constraint\footnote{Since the deviation in $U$ parameter is suppressed in this model and the uncertainty in $S$ and $T$ parameters will be reduced considerably, we set $U=0$.} \cite{Peskin:1990zt,Peskin:1991sw} both excluding and including recent controversial CDF result \cite{CDF:2022hxs}.
More concretely we use 
\begin{align}
S=0.00\pm0.07,~~T=0.05\pm0.06,
\label{Eq:2021_ST}
\end{align}
with the correlation of $\rho=0.92$ \cite{PDG2020} (denoted as 2021\,fit)
and 
\begin{align}
S=0.086\pm0.077,~~T=0.177\pm0.070,
\label{Eq:2023_ST}
\end{align}
with the correlation of $\rho=0.89$ based on the global fit \cite{deBlas:2022hdk} (denoted as 2023\,fit).
Fig.\,\ref{Fig:chi2_ST} shows $\chi^2$ of $S$ and $T$ parameters as a function of $m_\phi$ where $m_{H^+}=$150\,GeV (blue), 200\,GeV (orange) and 250\,GeV (green) is fixed.
Dashed and solid lines are drawn based on 2021\,fit and and 2023\,fit.
We see that the favored $m_\phi$ is different depending on the fit data.
For $m_{H^+}=150$\,GeV, 2023\,fit disfavors $m_t\le m_\phi\le200\,\GeV$ more than $2\,\sigma$, while 2021\,fit allows the mass window.

In short section summary, for the simultaneous explanation we need to set $m_t\le m_\phi\le200\,\GeV$ or $\mathcal{O}$(1)\,GeV level mass degeneracy among neutral scalars.

\section{Exotic top processes}
\label{Sec:ETP}
\begin{figure}[t]
\begin{center}
\includegraphics[scale=0.28]{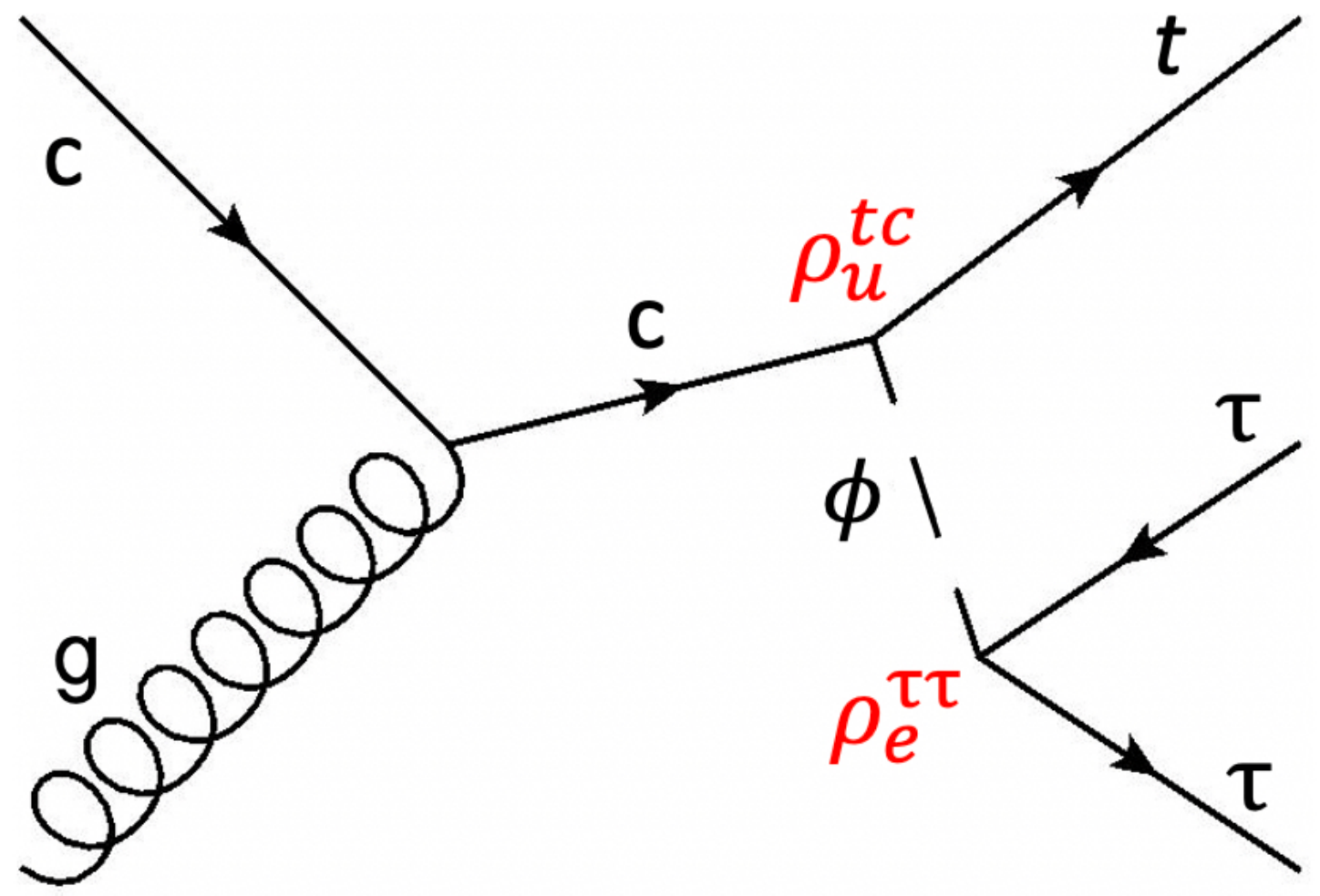}
\caption{
\label{Fig:dia_exoTics}
Representative diagram for $gc\to c\to t\phi\to t \tau\ov\tau$.
} 
\end{center}
\vspace{-.4cm}
\end{figure}

\begin{figure}[t]
\begin{center}
\includegraphics[scale=0.35]{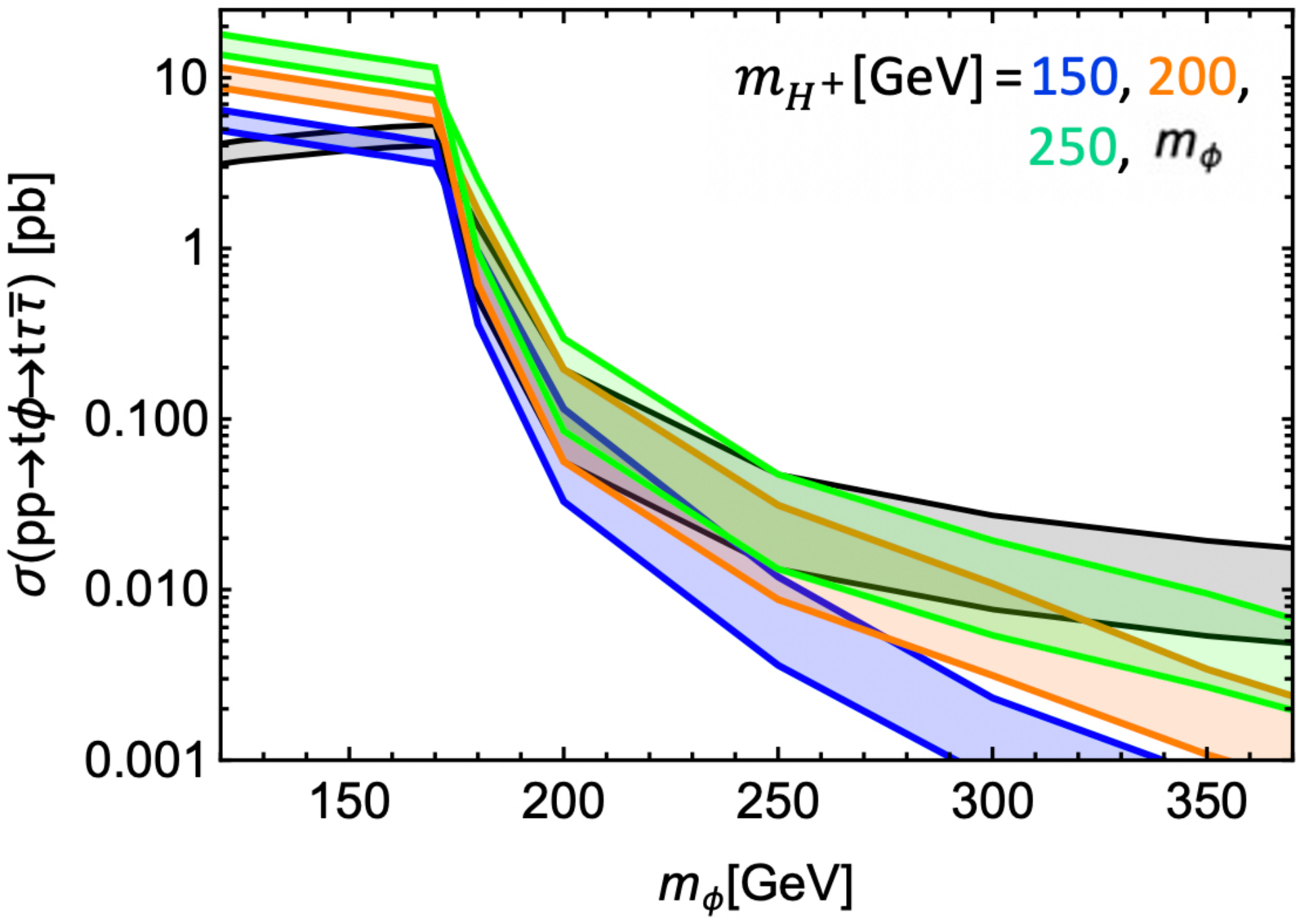}
\caption{
\label{Fig:pred_t2ta}
Prediction of $\sigma(pp\to t\phi \to t \tau\ov\tau)$ [pb] as a function of $m_\phi$ for the simultaneous explanation of deviations in $b\to s \ell\ov \ell$ and $b\to c\tau \ov\nu$. 
} 
\end{center}
\vspace{-.45cm}
\end{figure}

In order to fully probe the remaining mass window of $m_\phi$ we propose another top-associated process, namely $gc\to c\to t\phi\to t \tau\ov\tau$ where the relevant diagram is shown in Fig.\,\ref{Fig:dia_exoTics}.\footnote{It would be worthwhile to mention that $t\ov{t}$ inclusive cross section measurement still has an uncertainty of 70\,pb \cite{ATLAS:2020aln} and does not exclude the scenario with $gc\to c\to t\phi\to t \ov {t} c$ channel.}
In the mass window, even with the hierarchical coupling structure, BR($\phi\to\tau\ov\tau$) could be sizable due to the phase space suppression in $\phi\to t c$ decay.
The production cross section is calculated using {\sc\small MadGraph}5\_a{\sc\small MC}@{\sc\small NLO}~\cite{Alwall:2014hca} using {\sc\small NNPDF}2.3 \cite{Ball:2012cx} at the leading order in the five flavor scheme with $\sqrt{s}=13$\,TeV.
Fig.\,\ref{Fig:pred_t2ta} shows the cross section in pb as a function of $m_\phi$.
The prediction of the $1\,\sigma$ simultaneous explanation was obtained by fixing the charged Higgs mass $m_{H^+}=150$\,GeV (blue), 200\,GeV (orange), 250\,GeV (green) and $m_\phi$ (black).
It is observed that bands are overlapping and the cross section is as large as 30\,fb$\sim$10\,pb for the mass window.\footnote{For the numerical analysis we include $\phi\to H^\pm W^\mp$ if the phase space is available.}
A heavier charged scalar predicts the larger signal rate since it requires larger couplings. 

Estimating the size of the electroweak SM back ground (BG) is not difficult even for our mass range. 
For instance, $tZq$ and $thq$ production contribute to $t+\tau\ov{\tau}+q$ final state with cross section of $\simeq 50$\,fb  \cite{CMS:2018sgc} and $\simeq 5$\,fb \cite{LHCHiggsCrossSectionWorkingGroup:2016ypw} where $\tau\ov{\tau}$ comes from $Z$ and $h$ decay for each.
Therefore the contribution from those processes are expected to be moderate.
On the other hand, it is not easy to estimate the precise amount of the miss-tag associated BG {\it{e.g.}} from  $tW^-q\to t \tau \ov\nu+\slashed{j}$ and $t\ov{t}\to t W^-j \to t \tau \ov\nu+\slashed{j}$
where slashed final state will be miss-tagged as a hadronically decaying $\tau$ ($\tau_h$).
For the precise determination we need a considerable help from the experimental side and thus investigating the sensitivity of this channel is beyond the scope of this letter.\footnote{The charge asymmetry of the top quark would help to improve the sensitivity since the SM single top has the production asymmetry, while our signal does not have this feature.}
Actually Ref.\,\cite{CMS:2020mpn} searched for the $thq$ production with $h\to\tau\ov \tau$ with Run\,2 full data.
They set the upper limit of $\mu=8.1^{+8.2}_{-7.5}$ where $\mu$ denotes a signal strength.
This approximately leads to the upper limit on $\sigma(thq\to t\tau \ov\tau q )\lesssim 100\,$fb for $m_{\tau\tau}=125\,\GeV$.
Since the invariant mass of our signal is larger, the corresponding SMBG would be smaller and thus we can expect the better sensitivity.

\section{Summary and discussion}
\label{Sec:Summary}
Recently the charged Higgs solution to B anomalies became more interesting than ever.
The charged Higgs need to interact with left-handed bottom quark and thus can be a part of an additional doublet.
Hence a two Higgs doublet model is a minimal model and there are also two additional neutral scalars.
The Yukawa interaction of those scalars are related by SU(2)$_L$ rotation and the simultaneous explanation predicts distinctive signal at the LHC.
The theoretical proposals to probe the solution via charged Higgs mediated processes was made last year, however, the crucial process has not been tested experimentally yet.
Although, in the meantime, the ATLAS experiment reported the game changing constraint on the neutral scalars.
In this letter we reinterpret the ATLAS constraint and obtained the condition for the mass spectrum of the additional neutral scalars:
${\mathcal{O}}(1)\,\GeV$ mass degeneracy among $H$ and $A$ or $m_t\le m_\phi \le 200\,\GeV $ where $\phi$ denotes $H$ and $A$.
We also pointed out that the signal cross section of $gc\to t\phi \to t \tau\ov\tau$ could be as large as 10\,fb$\sim$10\,pb for the mass window.

Imposing a U(1) Peccei-Quinn symmetry \cite{Peccei:1977hh}, $\{H_1,\,H_2\} \to \{H_1,\,H_2 e^{i\alpha}\}$ can prohibit $\lambda_5$ and realize the mass degeneracy of additional neutral scalars \cite{Branco:2011iw}.
Although this symmetry should be broken since we also need Yukawa couplings, $\rho_u^{tc}$ and $\rho_e^{\tau\tau}$ and therefore the more complicated setup is necessary \cite{Ferreira:2010ir,Serodio:2013gka,Chiang:2015cba,Bjorkeroth:2018ipq}.

In general, other couplings $\it{e.g.}$ di-bottom quark coupling, namely $\rho_d^{bb}$ would be non-negligible.
For instance, one would think that $\mathcal{O}(10^{-2})$ of $\rho_d^{bb}$ could reduce the branching ratio of $\phi\to\tau\ov\tau$ thanks to the color factor and revive the scenario with $m_\phi\le m_t$.\footnote{It is noted that even in that case $C_{S_R}^\tau$, where the chirality of quarks are flipped in Eq.\,(\ref{Eq:Hamiltonian_RD}), can not be large to affect $R_{D^{(*)}}$ due to the $V_{cb}$ suppression \cite{Iguro:2017ysu}.}
Although this is difficult since the ATLAS collaboration searched additional particles in flavor changing top decays set $\mathcal{O}(10^{-4})$ upper bound on BR$(t\to q X) \times$BR$(X\to b\ov b)$ very recently \cite{ATLAS:2023mcc}.
Therefore an additional coupling to bottom quarks, does not save the scenario.
Since $c\to b$ miss tagging rate, $\epsilon_{c\to b}$ is about $15\sim20\,\%$ \cite{ATLAS:2022qxm}, even if neutral scalars decay into charm quarks, the scenario is difficult to survive the constraint.
On the other hand, $\rho_d^{bb}$ would be able to reduce signal rate of $gc\to t \tau\ov\tau$ process.

It would to worthwhile to emphasize that the ATLAS bound \cite{ATLAS:2022xpz} does not necessarily kill the solo $R_{D^{(*)}}$ solution even without mass degeneracy.
This is because that the contribution to $C_{S_L}$ is proportional to the coupling product of $\rho_u^{tc*}\rho_e^{\tau\tau}$ (see, Eq.\,(\ref{Eq:RD_simplified})) and hence the larger $\rho_e^{\tau\tau}$ allows the smaller $\rho_u^{tc}$.
If we want to avoid the ATLAS bound on $\rho_u^{tc}$ by setting $m_A,\,m_H\le 200\,\GeV$ instead, electroweak precision parameters at $2\,\sigma$ give the upper limit on the charged Higgs mass as $m_{H^+}\le 270\,\GeV$ ($290\,\GeV$) for Eq.\,(\ref{Eq:2021_ST}) (Eq.\,(\ref{Eq:2023_ST})). 
In this case, $t+\tau\ov\tau$ would provide a key test since BR$(\phi\to\tau\ov\tau)$ will be amplified compared to the scenario for the simultaneous explanation.

\section*{Acknowledgements}
I would like to thank Ulrich Nierste, Marco Fedele, Hiroyasu Yonaha and Teppei Kitahara for the useful discussion and great encouragement.
I appreciate Masaya Kohda for the exchange on the same-sign top signal in 2018.
I also appreciate Javier Montejo Berlingen, Tamara Vazquez Schroeder and Shigeki Hirose for the detailed information of Ref.\,\cite{ATLAS:2022xpz}.
The work is supported by the Deutsche Forschungsgemeinschaft (DFG, German Research Foundation) under grant 396021762-TRR\,257.

\bibliography{ref}

\end{document}